# Mn $L_{3,2}$ X-ray Absorption Spectroscopy And Magnetic Circular Dichroism In Ferromagnetic $Ga_{1-x}Mn_xP$

P.R. Stone[1,2], M.A. Scarpulla[1,2], R. Farshchi[1,2], I.D. Sharp[1,2], J.W. Beeman[2], K.M. Yu[2], E. Arenholz[3], J. Denlinger[3], E.E. Haller[1,2], and O.D. Dubon[1,2]

[1] *Department of Materials Science and Engineering, University of California, Berkeley, CA 94720*
[2] *Lawrence Berkeley National Laboratory, Berkeley, CA 94720*
[3] *Advanced Light Source, Lawrence Berkeley National Laboratory, Berkeley, CA 94720*

**Abstract.** We have measured the X-ray absorption (XAS) and X-ray magnetic circular dichroism (XMCD) at the Mn $L_{3,2}$ edges in ferromagnetic $Ga_{1-x}Mn_xP$ films for $0.018 \leq x \leq 0.042$. Large XMCD asymmetries at the $L_3$ edge indicate significant spin-polarization of the density of states at the Fermi energy. The spectral shapes of the XAS and XMCD are nearly identical with those for $Ga_{1-x}Mn_xAs$ indicating that the hybridization of Mn *d* states and anion *p* states is similar in the two materials. Finally, compensation with sulfur donors not only lowers the ferromagnetic Curie temperature but also reduces the spin polarization of the hole states.

## INTRODUCTION

The discovery that III-V semiconductors exhibit ferromagnetism when doped with a few atomic percent of Mn has led to unique possibilities for combined non-volatile information storage and processing [1]. Inter-Mn exchange is mediated in these ferromagnetic semiconductors (FMSs) by holes provided by substitutional manganese acceptors. We recently demonstrated the synthesis of a carrier-mediated ferromagnetic phase in $Ga_{1-x}Mn_xP$ using ion implantation and pulsed-laser melting (II-PLM) [2]. Unlike the holes in the prototypical FMS $Ga_{1-x}Mn_xAs$ where ferromagnetism is mediated by itinerant valence band holes, in $Ga_{1-x}Mn_xP$ these holes are localized in a Mn-derived band. $Ga_{1-x}Mn_xP$ is thus an important medium for probing the interplay between electronic structure, localization and carrier-mediated exchange. Here we report X-ray absorption spectroscopy (XAS) and X-ray magnetic circular dichroism (XMCD) measurements of ferromagnetic $Ga_{1-x}Mn_xP$.

## EXPERIMENTAL PROCEDURE

GaP (001) wafers were implanted with 50 keV $Mn^+$ to doses between $4.5 \times 10^{15}$ and $2.0 \times 10^{16}$ /cm$^2$. Samples for compensation studies were subsequently implanted with 60 keV $S^+$ to doses between $1.0 \times 10^{15}$ and $7.3 \times 10^{15}$ /cm$^2$. Samples were irradiated in air with a single 0.44 J/cm$^2$ pulse from a KrF ($\lambda = 248$ nm) excimer laser having FWHM of 18 ns, and subsequently etched in concentrated HCl for 24 hours. The concentrations of substitutional manganese ($Mn_{Ga}$) were determined by the combination of SIMS and ion beam analysis. We define x as the peak $Mn_{Ga}$ concentration as we have shown that the magnetic properties of II-PLM films are dominated by the film region having maximum x [3]. DC magnetization was measured by SQUID magnetometry along <110> in-plane directions. Room temperature XAS was performed at beamline 8.0 at the Advanced Light Source (ALS). Low-temperature XAS and XMCD measurements were carried out at beamline 4.0.2 at the ALS in applied fields of ±5.4 kOe. Data were collected with the field and beam oriented 30º from the plane of the samples along a <110> in-plane direction with 90% circular polarization of the incident X-rays.

## RESULTS AND DISCUSSION

The main panel of Fig. 1 presents Mn $L_{3,2}$ TEY XAS spectra taken at 17 K with the field and photon helicity parallel ($I^+$) and antiparallel ($I^-$) for a sample having x=0.034. The XMCD ($I^+-I^-$) spectrum is also shown. Room temperature XAS demonstrated that these spectra arose from well-incorporated $Mn_{Ga}$ and not from surface oxide phases which obscured early XMCD studies of $Ga_{1-x}Mn_xAs$ [4]. Strong XMCD is present at both the $L_3$ and $L_2$ edges indicating a large spin polarization of states derived from Mn $d$ levels at $E_F$. While TEY mode probes depths of under 10 nm the total fluorescence yield (TFY) mode can probe depths on the order of tens of nanometers. Since the $Mn_{Ga}$ peak in these films occurs 20-30 nm from the surface, TFY is a better probe of their bulk magnetic properties. The inset of Fig. 1 compares the magnitude of the asymmetry $((I^+-I^-)/(I^++I^-))$ for the $L_3$ edge at 17 K as measured in TEY and TFY modes. When corrected for incident angle and photon polarization, the TFY data exhibit a maximum asymmetry value of 0.70±0.04 in all samples except for the one having x=0.018. This is primarily because the $T_C$ of 18 K of this film is very close to the measurement temperature. The TEY data are generally lower than the TFY data, which is consistent with lower Mn concentration and magnetic coupling near the surface of the films.

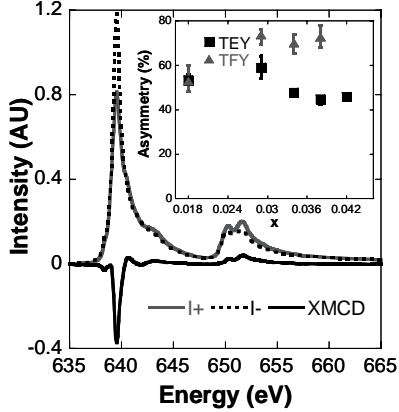

**FIGURE 1.** (main) Mn $L_{3,2}$ TEY XAS spectra for magnetization and helicity parallel ($I^+$) and antiparallel ($I^-$) as well as the difference (XMCD) spectrum for a $Ga_{1-x}Mn_xP$ sample with x=0.034 measured at 17 K. (inset) TEY and TFY asymmetry at the Mn $L_3$ peak versus x.

The similarity between the XAS and XMCD lineshapes for $Ga_{1-x}Mn_xP$ and those reported for $Ga_{1-x}Mn_xAs$ [4,5] is remarkable. Because XAS and XMCD lineshapes are strongly influenced by the hybridization of the $t_2$-symmetric Mn $d$ orbitals with the neighboring anion $p$ orbitals, this suggests that the bonding and $p$-$d$ exchange between Mn and As or P in dilute alloys are substantially similar.

The data in Table 1 show a marked decrease in $T_c$ with increasing S concentration. SIMS and channeling RBS yielded x≈ 0.041 as well as a Mn substitutionality of 85-90% for all S-doped films, indicating that Mn incorporation is not significantly affected by S concentration. Indeed it is likely that S enhances Mn incorporation (by as much as 15% in this study). Thus, the decrease in $T_c$ is attributed to a decrease in hole concentration ($p$) due to the presence of S donors. Furthermore, we find that the XMCD decreases monotonically with S concentration despite the $T_c$ of most films being well above the measurement temperature (Table 1). This result indicates that inter-Mn exchange is intimately related to hole concentration. While changes in $Mn_{Ga}$ and $p$ both affect $T_c$, only modulation of $p$ significantly affects the spin polarization of carriers for $T>T_c$.

**TABLE 1.** Properties of $Ga_{0.959}Mn_{0.041}P$:S

| Sulfur Implant Dose ($S^+/cm^2$) | $T_c$(K) | TFY XMCD Asymmetry at 17 K and ±5 kOe(%) |
|---|---|---|
| 0 | 56[*] | 70±4[$] |
| $1.0×10^{15}$ | 39 | 55±3 |
| $2.5×10^{15}$ | 35 | 48±3 |
| $5.0×10^{15}$ | 29 | 37±2 |
| $7.3×10^{15}$ | 21 | 21±2 |

[*]Interpolated from Ref. 6.  [$]Interpolated from Fig. 1

## CONCLUSION

X-ray absorption studies of ferromagnetic $Ga_{1-x}Mn_xP$ have been reported as a function of $Mn_{Ga}$ and hole concentrations. Large XMCD is observed for 0.018≤x≤0.042, which is consistent with, but not exclusive to, a spin-polarized impurity band. The XMCD and XAS lineshapes are nearly identical to those observed in $Ga_{1-x}Mn_xAs$ suggesting similar electronic environments around $Mn_{Ga}$ in the two systems despite differences in carrier localization. Both $T_c$ and XMCD decrease monotonically with increasing S donor concentration, a hallmark of a hole-mediated ferromagnetic phase.

## ACKNOWLEDGMENTS

This work is supported by the Director, Office of Science, Office of Basic Energy Sciences, Division of Materials Sciences and Engineering, of the U.S. Department of Energy under Contract No. DE-AC02-05CH11231. PRS acknowledges support from a NDSEG Fellowship.